# Hybrid Movie Recommender System based on Resource Allocation

Mostafa Khalaji[1]    Chitra Dadkhah[1]    Joobin Gharibshah[2]

[1]Faculty of Computer Engineering, K. N. Toosi University of Technology, Tehran, Iran
[2]University of California – Riverside, CA

## Abstract

Recommender Systems are inevitable to personalize user's experiences on the Internet. They are using different approaches to recommend the Top-K items to users according to their preferences. Nowadays recommender systems have become one of the most important parts of large-scale data mining techniques. In this paper, we propose a Hybrid Movie Recommender System (HMRS) based on Resource Allocation to improve the accuracy of recommendation and solve the cold start problem for a new movie. HMRS-RA uses a self-organizing mapping neural network to clustering the users into N clusters. The users' preferences are different according to their age and gender, therefore HMRS-RA is a combination of a Content-Based Method for solving the cold start problem for a new movie and a Collaborative Filtering model besides the demographic information of users. The experimental results based on the MovieLens dataset show that the HMRS-RA increases the accuracy of recommendation compared to the state-of-art and similar works.

## 1. Introduction

Facing the huge amount of internet information and increasing online shopping, the need for recommendation systems became obvious as a way to guide users toward their preferences. In the recommender systems, users are the set of the actors in the system who have interactions with each other and items are the set of entities that can be liked or observed by users [1]. In recommender systems, when the number of users or items grows exponentially at an enormous speed, scalability becomes a problem, so a recommendation method should be fast and efficient on a small dataset as well as a large one. On the other hand, lacking the information of new users or items causes the cold-start problem which affects the accuracy of prediction. One of the techniques to handle this problem was introduced by Liben-Nowell and Kleinberg in 2007 [2,3] which find a similarity between two entities in a network.

There are three foundation approaches in recommender systems [1]: 1: Collaborating Filtering (CF) which the recommendation is based on the performance or behavior of the other users in the system. 2: Content-Based Method (CBM) in which the recommendation is based on active user history and the descriptive features of items. 3: Knowledge-Based Systems (KBS), in which the recommendation is based on explicit knowledge regarding users' requirements. Finally, Hybrid Systems (HS) are a combination of different approaches to improve the performance of the system [1,4].

Meta heuristic-based recommender systems aim to find a possible solution for complex optimization tasks. In these systems, a set of possible solutions search through the solution space to find the optimal region of the search space and discover a near-optimal solution in a reasonable time. For achieving this aim, a set of search operators are designed in such a way that the most successful solutions are let in to be evolved through the next iterations.

Resource Allocation (RA) is a measure used to compute the closeness of nodes based on their shared neighbors. The Resource Allocation algorithm was introduced in 2009 by Tao Zhou et al as part of a study to predict links in various networks [5].

According to the RA algorithm, a pair of nodes—which are not connected directly—can share their resources through their common neighbors. RA-reliability indicates the degree of reliability for the estimated similarity between the users $u_i$ and $u_j$ [6]. RA is used to find out missing relations or possible future relations in the network [4], e.g., recommend friends on social networks [7] or recommend an item in online shopping [8].

The neural network has the advantages of strong feature extraction, effective learning, and simple process. Self-Organizing Map (SOM) is an unsupervised neural network and it is more suitable for the development of complex and different recommender systems. We could cluster similar users or items in the same feature space area using SOM neural network [9].

In this paper, we propose a hybrid Movie Recommender System based on Resource Allocation (HMRS-RA) to handle scalability, cold-start problems, and increases the accuracy of recommendation.

The novelty of our approach is the use of RA for computing the weights for user similarity in person similarity measure to increase the precision of the predictions.

The structure of this paper is organized as follows: Section 2 presents the summary of various recent research. Section 3 describes the structure of HMRS-RA. In Section 4, we demonstrate the results of the implementation and evaluation of HMRS-RA. Finally, the conclusion is presented in Section 5.

## 2. Related Work

As mentioned in the abstract section, HMRS-RA combines the CF and CBM to improve the efficiency of recommender systems, so we have studied the researches in three categories: collaborative filtering RS, content-based RS, and hybrid RS. Because most researches in CFRS focus on Metaheuristic, we have also explained these researches. We have summarized the researches in Fig.1.

### 2.1. CF approach

We divided the researches in CFRS into probabilistic and trust approaches. In the probabilistic approach, Ma et al. proposed a factor analysis approach that is called Sorec, which is based on probabilistic matrix factorization to solve the data sparsity by employing both users' social network information and rating records [10]. Chaney et al. developed Social Poisson Factorization (SPF), a probabilistic model that incorporates social network information into a traditional factorization method; SPF introduces the social aspect to algorithmic recommendation [11]. Chen et al. proposed a probabilistic recommender system that uses the clustering method. Their system learns co-preference patterns from historical transaction data and recommends items accordingly [12].

In the trust approach, we divided the researches into matrix factorization and link prediction methods. In the matrix factorization method, Ma et al. proposed a novel probabilistic factor analysis framework, which naturally fuses the users' tastes and their trusted friends' favors together [13]. Zhang et al. developed a novel healthcare recommendation system called iDoctor, which is based on hybrid matrix factorization methods. iDoctor differs from previous work in the following aspects: (1) emotional offset of user reviews can be unveiled by sentiment analysis and can be utilized to revise original user ratings; (2) user preference and doctor features are extracted by Latent Dirichlet Allocation and incorporated into conventional matrix factorization [14]. Jamali et al. proposed a model-based approach for the recommendation in social networks, employing matrix factorization techniques and incorporated the mechanism of trust propagation into the model. Trust propagation is a crucial phenomenon in the social sciences, social network analysis and trust-based recommendation [15]. Li et al. proposed a novel recommendation method called TruCom. In a multi-category item recommendation domain, TruCom, first generates a domain-specific trust network pertaining to each domain and then builds a unified objective function for improving recommendation accuracy by incorporating the hybrid information of direct and indirect trust into a matrix factorization recommendation model [16]. Guo et al. proposed TrustSVD, which is a trust-based matrix factorization technique. By analyzing the social trust data from four real-world data sets, they have concluded that not only the explicit rating but also the implicit influence of trust should be considered in a recommendation model [17]. Al Hasan et al. gathered some representative link prediction methods according to the type of the models. They have considered three types of models: 1: the traditional (non-Bayesian) model which extracts a set of features to train a binary classification model. 2: the probabilistic model which models the joint-probability among the entities in a network using Bayesian graphical models. 3: the linear algebraic model computes the similarity between the nodes in a network by rank-reduced similarity matrices [18]. Xia et al. proposed a recommendation algorithm Improved Weighted Network-Based Inference (INBIw) that improves the original weighted network-based inference by introducing a tunable parameter $\beta$ to reduce the influence of high-degree nodes. In order to evaluate the recommendation performance of INBIw, ranking position rate and hitting rate are calculated [19]. Zhao et al. proposed a method for supporting resource allocation in business process management. In their method, resource allocation is considered as a multi-criteria decision problem and solved by a new entropy-based clustering ensemble approach. By mining resource characteristics and task preference patterns from past process executions, the right resources could be recommended to improve resource utility [20]. Lu et al. summarize link prediction algorithms, emphasizing the contributions from physical perspectives and approaches, such as the random-walk-based methods and the maximum likelihood methods. They also introduced three typical applications: reconstruction of networks, evaluation of network evolving mechanism, and classification of partially labeled networks [21]. Fangyi Hu introduced a three-segment similarity measure method for the collaborative filtering model. He improved the performance of similarity measure by computing the similarity between users based on the number of user ratings along with item similarity and user attribute similarity [22].

### 2.1.1. Clustering approach

In this section, we have explained the methods used in the papers that focus on clustering in model-based CF. Belacel et al. introduced a scalable recommender system based on a collaborative filtering approach. They improved the time and accuracy of their proposed system using the split-merge clustering algorithm [23]. Kant et al. introduced a method to determine the selection of the center of the cluster in the K-means clustering operation. Their method was able to solve the data sparsity problem [24]. Wang et al. introduced a new method called the CDIE. They used the Co-Clustering method to learn cross-domain comprehensive representations of items by collectively leveraging single-domain and cross-domain sessions within a unified framework. Their method solved the data sparsity problem [25]. Rafiee et al. proposed a similarity-based link prediction algorithm, referred to as CNDD, which in this algorithm the similarity score is determined according to the structure and specific characteristics of the network, as well as the topological characteristics. In their proposed method, a new metric for link prediction is introduced, considering clustering coefficient as a structural property of the network. Moreover, their method also considers the neighbors of shared neighbors of each pair of nodes, which leads to achieving better performance than the other similar link prediction methods [26]. Zhu et al. proposed link prediction indices based on both Network Structure and Topic Distribution (NSTD). Their approach makes full use of the network characteristics, such as homophily, transitivity, clustering, and degree heterogeneity. And they combined these characteristics with topic similarity when constructing indices based on both directly and indirectly connected nodes [27]. Liu et al. introduced a novel Collaborative Linear Manifold Learning (CLML) algorithm which can optimize the consistency of nodes similarities using the manifolds embedded between the target and the auxiliary network [28]. Mazzouzi et al. proposed a new effective recommender system for TED (Technology, Entertainment, and Design) talks that first groups the users according to their preferences and then provides a powerful mechanism to improve the quality of recommendations for users. In their system, the authors used the Pearson Correlation Coefficient (PCC) method and TED talks to create the TED user-user matrix. Then, they used the k-means clustering method to group the same users in clusters and create a predictive model. Finally, they used this model to make relevant recommendations to other users [29]. Xiaopan et al. for solving the data sparsity problem in CF, proposed a SOM clustering collaborative filtering algorithm based on Singular Value Decomposition (SVD) which reduces the dimensions of the original matrix. by decomposing to the item and user latent factor [30].

Parvin et al. proposed a novel CF method for predicting missing ratings accurately. Their proposed method, called TCFACO (Trust-aware Collaborative Filtering Ant Colony Optimization), used trust statements as a rich side information with Ant Colony Optimization (ACO) method [31]. For increasing the accuracy of the recommendation of user-based, Tohidi et al. proposed a hybrid approach based on clustering and evolutionary algorithm. They combined the K-means clustering method along with two Metaheuristic algorithms such as FOA and APSO [32]. Khodaverdi et al. proposed a movie hybrid recommender system based on clustering and popularity. Their system clusters the users who were similar to each other by using the K-means clustering method and uses rating popularity to predict the users' preferences for specific movies [33]. Khalaji proposed a new recommender system called NWS_RS for movie recommendations. His method was able to personalize the recommendation by segmenting users' age. NWS_RS used the New Weighted Similarity (NWS) for improving the accuracy of prediction of unobserved movies for active users. NWS_RS managed the scalability problem and solved the data sparsity problem [34]. Khalaji et al proposed a new recommender system called CUPCF which was a combination of two similarity measures in CF to solve the data sparsity and better recommendation. CUPCF used two similarity measures simultaneously as a new method for decreasing the error rate of the system [35].

## 2.2. Content-based approach

In content-based methods, Mooney et al. proposed a content-based book recommender system for text categorization which their approach has the advantage of being able to recommend previously unrated items to users with unique interests and to provide explanations for its recommendations [36]. Deldjoo et al. proposed a new content-based recommender system that encompasses a technique to automatically analyze video contents and to extract a set of representative stylistic features (lighting, color, and motion) grounded on existing approaches of Applied Media Theory [37]. Van den Oord et al. use a latent factor model for the recommendation, and predict the latent factors from music audio when they cannot be obtained from usage data [38]. Yang et al. proposed a movie recommendation system according to scores that the users have provided. In view of the movie evaluation system, the impacts of access control and multimedia security are analyzed, and secure hybrid cloud storage architecture is presented. Mobile-Edge Computing (MEC) technology is used in the public cloud which guarantees the high-efficiency requirements of the transmission of the multimedia content. The processes of the system include registration, user login, role assignment, data encryption, and data decryption [39]. Wang et al. proposed a content-based recommender system for computer science publications. Their system recommends suitable journals or conferences with a priority order based on the abstract of a manuscript. To follow the fast development of computer science and technology, a web crawler is employed to continuously update the training set and the learning model. To achieve the interactive online response, they propose an efficient hybrid model based on chi-square feature selection and softmax regression [40]. Rahimpour et al. introduced a new method for a content-based filtering recommender system. They use the interactions of each user and analyze them to propose a new user model and capture user's interests. Their system built the user model based on a Bayesian framework called the Dirichlet Process Mixture Model. They improved the accuracy of their system in comparison to other methods [41].

## 2.3. Hybrid approach

In hybrid systems, Lee et al. proposed a new recommender system that combines collaborative filtering with Self-Organizing Map (SOM) neural network. First, all users are segmented by demographic characteristics and users in each segment are clustered according to the preference of items using the SOM neural network [42]. Nadi et al. proposed a fuzzy recommender system (FARS) based on the collaborative behavior of ants. FARS works in two phases: modeling and recommendation. First, user's behaviors are modeled offline and the results are used in the second phase for the online recommendation. Fuzzy techniques provide the possibility of capturing uncertainty among user interests and Ant-based algorithms optimize the solutions for predicting phase. The performance of FARS is evaluated using log files of Information and Communication Technology Center of Isfahan municipality in Iran and have compared with Ant-based Recommender System (ARS) [43]. Roh et al. proposed a three-step CF recommendation model, which is composed of profiling, inferring, and predicting steps while considering prediction accuracy and computing speed simultaneously. Their model combines a CF algorithm with two machine learning methods, Self-Organizing Map (SOM) and Case-Based Reasoning (CBR) by changing an unsupervised clustering problem into a supervised user preference reasoning problem, which is a novel approach for the CF recommendation field [44]. May et al. proposed a neural networks-based clustering collaborative filtering algorithm in the e-commerce recommendation system. Their algorithm tries to establish a classifier model based on Back Propagation (BP) neural network for the pre-classification of items. They analyzed and discussed their algorithm from multiple aspects [45]. Kim et al. proposed a robust document context-aware hybrid method, which integrates Convolutional Neural Network (CNN) into Probabilistic Matrix Factorization (PMF). Their method captured contextual information using the statistics of items [46]. Katarya et al. proposed a component of Hybrid Music Recommender Systems (HMRS), which combined context-sensitive and collaborative filtering approaches. Their method used the timestamp of user rating for modeling user behaviors. They used the Depth-First-Search (DFS) algorithm which traverses the whole graph through the paths in different contexts and generated the ranked list of recommended items using the Bellman-Ford algorithm with multi-layer context graph [47]. De Pessmier et al. proposed a recommender system that offers personalized recommendations for travel destinations to individuals and groups. These recommendations are based on the users' rating profile, personal interest, and specific demands for their next destination. Their recommendation algorithm was a hybrid approach that combined content-based collaborative filtering and knowledge-based models. For groups of users, such as families or friends, individual recommendations are suggested into group recommendations, with an additional opportunity for users to give feedback on these group recommendations. A group of test users has evaluated the recommender system using a prototype web application [48]. Wei et al. proposed a hybrid movie recommendation approach using tags and ratings. First, they constructed social movie networks and a preference-topic model. Then, they extracted, normalized, and reconditioned the social tags according to user preference based on social content annotation. Finally, they enhanced the recommendation model by using supplementary information based on user historical ratings [49]. Deldjoo et al. proposed multimedia recommender systems called MMRS. They combined content-based and collaborative filtering approaches. The target of their system was the recommendation of music, movies and images using deep learning and feature extraction [50]. Tarus et al. proposed a new hybrid recommender system for e-learning. Their system used sequential pattern mining called SPM along with context-awareness and collaborative filtering approach for suggesting learning resources to the users. They improved the quality and accuracy of their system [51].

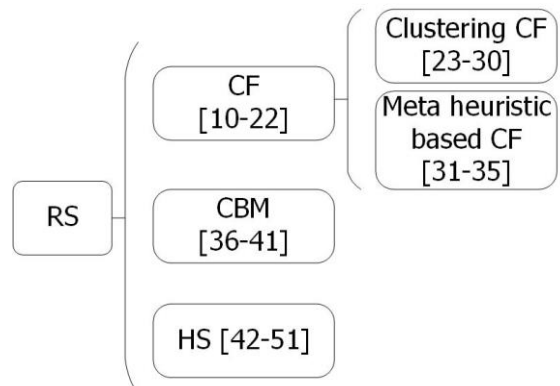

**Fig. 1:** The classification of related work

# 3. Proposed HMRS-RA

HMRS-RA consists of two phases: online and offline. In the offline phase, the preprocessing of the data is done and then the recommendation is made on the online phase as is depicted in Fig. 2.

In the offline phase, HMRS-RA filters users based on their gender (feminine and masculine) and age (range of 20-39 years and range of 40-60 years). In this step, four rating matrices ($R_{male}$, $R_{female}$, $R_{age(20-39)}$, $R_{age(40-60)}$) are generated according to the original rating matrix, which consists of the rating of each user for observed movies. Where, $R_{male}$ indicates groups of male and $R_{age(40-60)}$ represents the rating of the users who are older than 40 years old and below 60. The ratings are from the set $\{1, 2, 3, 4, 5\}$ in which a rating of 1 indicates an extreme dislike and a rating of 5 indicates the extreme like for a movie.

HMRS-RA identifies sets of similar users based on movies rating using the Self-Organizing Map (SOM) clustering method and uses N clusters to build up a model to predict the rating of unobserved movies for active users in the online phase. For each category of users, HMRS-RA defines the most preferred genres of movies rated by users. We consider the five popular genres of the movie, such as Action, Adventure, Comedy, Drama and Romance.

At the end of the offline phase, four rating matrices: $R'_{male_{c1}}$, $R'_{male_{c2}}$, $R'_{female_{c1}}$, $R'_{female_{c2}}$ with $m \times 5$ dimensions where m is the number of users and 5 is the number of genres will generate. Each entry of $R'_{male_{c1}}$ and $R'_{male_{c2}}$ is the average of the rating of the observed movies that belong to five popular genres for masculine users in Clusters.

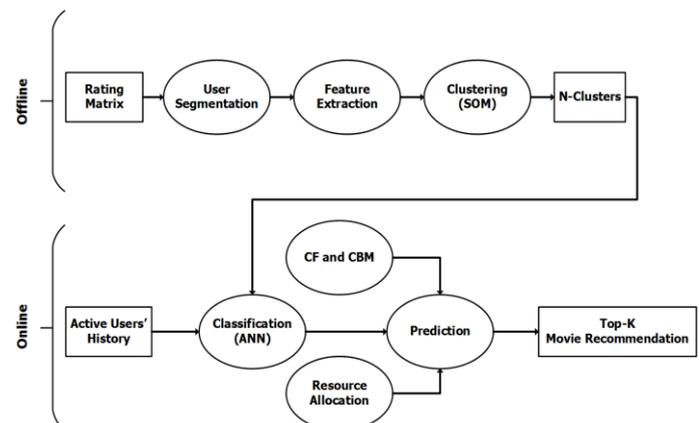

**Fig. 2:** The HMRS-RA structure

On the other side, another four rating matrices: $R'_{age(20-39)_{c1}}$ , $R'_{age(20-39)_{c2}}$ , $R'_{age(40-60)_{c1}}$ , $R'_{age(40-60)_{c2}}$, with $m \times 5$ dimensions where m is the number of users and 5 is the number of genres will generate. Each entry of $R'_{age(20-39)c1}$ and $R'_{age(20-39)c2}$ are the average of the rating of the observed movies that belong to five popular genres for users who are older than 19 years and below 40 in Clusters.

In the online phase, first HMRS-RA finds the cluster that the active user belongs to using a three layers Artificial Neural Network (ANN) classification and then detects preferred genre using CBM (Content-Based Method) and predicts the rating of unobserved movies for the active user using CF method. After determining the active user cluster, the similarity between the active user and the users in his/her cluster is calculated by Pearson similarity measure as shown in Eq. (1) to determine the k-neighbors of the active user.

$$Sim(u,v)$$
$$= \frac{\sum_{k \in I_u \cap I_v}(Rate_{uk} - \mu_u) \cdot (Rate_{vk} - \mu_v)}{\sqrt{\sum_{k \in I_u \cap I_v}(Rate_{uk} - \mu_u)^2} \cdot \sqrt{\sum_{k \in I_u \cap I_v}(Rate_{vk} - \mu_u)^2}} \quad (1)$$

Where K is the set of the same movies observed by user u and v. $Rate_{uk}$ indicates the rating of active user $u$ for movie $k$ and $\mu_u$ is the average movie rating given by user $u$. The value of $Sim(u,v)$ always lies in the range [-1, 1]. A value 1 indicates the most similarity between users while a value -1 indicates users are not similar. The similarity between two users who have rated the same unpopular movies is stronger than the similarity between users who have rated the same popular movies. Therefore HMRS-RA calculates the weight for the similarity between the users $u_i$ and $u_j$ using the Resource Allocation (RA) weighting method as shown in Eq. (2).

$$R_{RA}(u_i, u_j) = \sum_{z \in \Gamma(u_i) \cap \Gamma(u_j)} \frac{1}{k_z} \quad (2)$$

Where $k_z$ is the number of users that rated the movie z. $\Gamma(u_i)$ is the set of neighbors of user $u_i$ that rated movie z. So HMRS-RA predicts the rating of the unobserved movie $i$ for active user $u$ using Eq. (3).

$$Predict(u,i)$$
$$= \mu_u$$
$$+ \frac{\sum_{j=1}^{m}(r_{v_{j},i} - \mu_v).Sim(u,v_j).R_{RA}(u,v_j)}{\sum_{j=1}^{m}|Sim(u,v_j).R_{RA}(u,v_j)|} \quad (3)$$

Where $m$ is the number of neighbors for active user u and $\mu_u$ is an average of observed movies rating of user $u$ and $r_{v_{j},i}$ indicates the rating done by user $v_j$ for movie $i$. We consider half of the users in the active user's cluster for the value of m. After predicting the rating of all unobserved movies, HMRS-RA recommends the top K-movies to active user u by ranking the predicted rating values.

# 4. Evaluation of HMSRS-RA

The effectiveness of HMRS-RA is evaluated on MovieLens [52] data set which consists of 943 users, 1682 items with 100,000 user ratings for movies. The ratings are from the set {1, 2, 3, 4, 5} that indicates the level of like or dislike of the observed movies.

For evaluating the HMRS-RA, we use the five-fold cross-validation algorithm. The cross-validation procedure consists of 5 iterations and in each iteration, 80 % of the data and the rest of the data (20%) is considered as training and test data, respectively [53]. The initial weight values in the SOM clustering method consider randomly, so each fold is repeated 10 times with independent running. We calculated the average value of the Mean Absolute Error (MAE) using Eq. (4) over different iterations of cross-validation.

$$MAE = \frac{\sum_{i=1}^{n}|\hat{r}_{u,i} - r_{u,i}|}{n} \quad (4)$$

Where, $\hat{r}_{u,i}$ is the predicted rating value of movie $i$ by user $u$ with HMRS-RA, $r_{u,i}$ is the actual rating value given by user $u$ for movie $i$ and $n$ is the number of the predicted values.

In our experiment, first, we estimated the real rating value for each genre by calculating the average rating value of observed movies of each genre as shown in Table 1, for the active user. Then we predicted the rating of the genre that its real rating is more than value 4 by HMRS-RA and calculated the MAE as shown in Table 2. We considered the genre which has the highest and lowest MAE value as the worst case and the best case, respectively. We predicted the rating of movies of the worst and best case by HMRS-RA and calculate the MAE as shown in Table 3 based on the initial rating matrix. Finally, the overall MAEs for the worst and the best cases are calculated by multiplication of MAEs in Table 2 and Table 3 [54] as shown in Table 4.

**Table 1**: The genre ratings by an active user

| Genre | Adventure | Romance | Comedy | Drama | Action |
|-------|-----------|---------|--------|-------|--------|
| Rate  | 4         | 4       | 3      | 1     | 4      |

**Table 2**: The MAE value for preferred genres for an active user

| Genre | Adventure | Romance | Action |
|-------|-----------|---------|--------|
| MAE   | 0.399     | 0.520   | 0.481  |

**Table 3**: The MAE value of the worst and the best case for an active user

| Genre | Worst case | Best case |
|-------|------------|-----------|
| MAE   | 0.71       | 0.60      |

**Table 4**: Overall MAE of HMRS-RA for an active user

| Genre | Worst Case | Best Case |
|-------|------------|-----------|
| MAE   | 0.3692     | 0.2394    |

**Table 5**: The MAE of HMRS-RA after 50 iteration

| HMRS-RA | MAE | | | |
|---------|-----|-------|-------------|-------------|
|         | Men | Women | 20-39 years | 40-60 years |
| Best case | 0.16043 | 0.23653 | 0.16881 | 0.23584 |
| Worst case | 0.30656 | 0.45268 | 0.29701 | 0.44450 |
| overall | **0.233495** | **0.344605** | **0.23291** | **0.34017** |

Fig. 3 and Fig. 4 indicate the comparison of HMRS-RA with the CF-RA in [6] and the RS that combined the traditional CF and SOM methods according to the MAE criteria. The results show the efficiency of our proposed algorithm and the increasing accuracy of recommendation by HMRS-RA.

Table 6 and Table 7 indicate the comparison of HMRS-RA with the latest work in [22], [24] and all of the methods in recommender systems that have been cited in them on recommender systems according to the MAE criteria. This section focuses on various measures that are related to new user cold-start problems and they are Proximity-Impact-Popularity (PIP), NHSM (New heuristic similarity measure), and other methods for computing users similarity such as Pearson, Cosine. BCF is a similarity measure based on Bhattacharyya coefficient [55]. The three-segment similarity measure is a model for solving cold-start and data sparsity problems in the recommender system. K-means is a clustering method in recommender systems and K-Means Leader is a new clustering collaborative framework, which improves the quality of clustering and recommendations.

**Table 6**: The evaluation of algorithms according to user gender

| Method | MAE | |
|---|---|---|
| | Men | Women |
| K-Means Leader [24] | 0.74 | 0.74 |
| K-Means [24] | 0.755 | 0.755 |
| Three-Segment Similarity [22] | 0.75 | 0.75 |
| BCF [22] | 0.78 | 0.78 |
| NHSM [22] | 0.83 | 0.83 |
| PIP [22] | 0.86 | 0.86 |
| Cosine [22] | 0.865 | 0.865 |
| Pearson [22] | 0.87 | 0.87 |
| **HMRS-RA** | **0.233495** | **0.344605** |

**Table 7**: The evaluation of algorithms according to user age

| Method | MAE | |
|---|---|---|
| | Ages 20 to 39 | Ages 40 to 60 |
| K-Means Leader [24] | 0.74 | 0.74 |
| K-Means [24] | 0.755 | 0.755 |
| Three-Segment Similarity [22] | 0.75 | 0.75 |
| BCF [22] | 0.78 | 0.78 |
| NHSM [22] | 0.83 | 0.83 |
| PIP [22] | 0.86 | 0.86 |
| Cosine [22] | 0.865 | 0.865 |
| Pearson [22] | 0.87 | 0.87 |
| **HMRS-RA** | **0.23291** | **0.34017** |

# 5. Conclusion

In this paper, we proposed a Hybrid Movie Recommender System which combines collaborative filtering and content-based filtering to solve the cold-start problem for new items. By considering the contextual information such as genre, HMRS-RA would diminish the cold start problem for new movies according to their genre. The proposed method (HMRS-RA) solves the scalability problem using clustering to reduce the dimensionality of the data. By considering the resource allocation as a weight for detecting the similarity between users in each cluster, we improved the performance of the recommendation comparing with a number of state-of-the-art and latest work in recommender systems. The experimental results showed that the MAEs of our proposed algorithm are 0.23, 0.34, 0.23, and 0.34 for men, women, age of 20-39 and age of 40-60, respectively. So, HMRS-RA increased the accuracy of recommendation. In the future, we would like to classify the users based on deep learning approaches such as convolutional neural networks in the case of a large dataset.

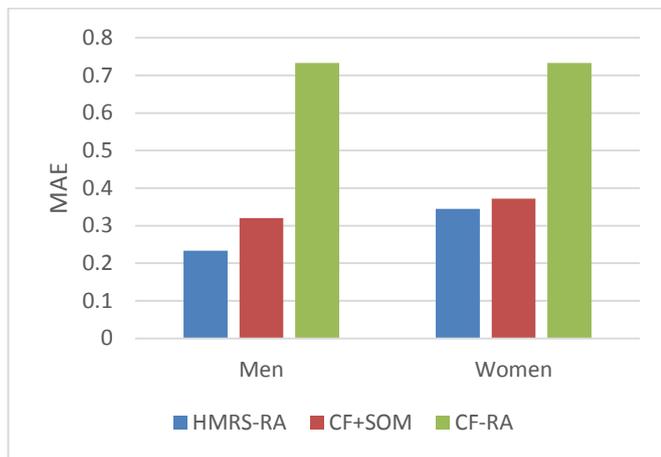

**Fig. 3**: The comparison of methods according to user gender

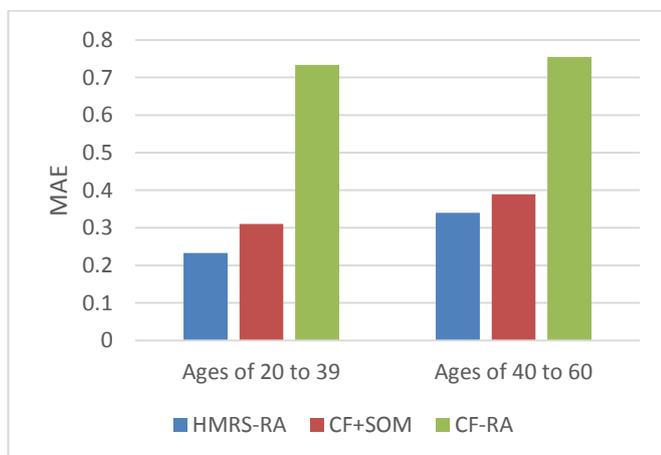

**Fig. 4**: The comparison of methods according to user age

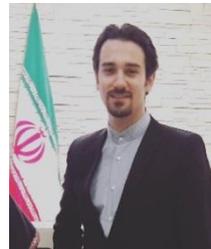

**Mostafa Khalaji** received the B.Sc. degree in computer engineering (Software) from Sadra Institute of Higher Education, Tehran, Iran, in 2015, and the M.Sc. degree in computer engineering (Artificial Intelligence) from K. N. Toosi University of Technology, Tehran, Iran, in 2017. He is a lecturer at Islamic Azad University, Shahr-e-Qods Branch and Sadra Institute of Higher Education. He is a member of IEEE and IAENG. His current research interests include Recommender Systems, Machine Learning, Social Network Analysis, and Data Mining.
Email: Khalaji@email.kntu.ac.ir

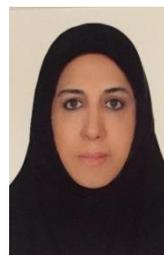

**Chitra Dadkhah** received the B.Sc. degree in software engineering from Shahid Beheshti University, Tehran, Iran, in 1990, the M. Sc. degree in computer engineering (Artificial Intelligence) from the IASI Department, University of Paris 11, Orsay, France, 1993. She also received the Ph.D. degree in computer engineering from the Department of Computer Engineering & IT, Amir Kabir University (Poly technique), Tehran, Iran, 2005. She is an Assistant Professor in the Computer Engineering Faculty of K. N. Toosi University of Technology, Tehran, Iran. Her research interests include Evolutionary & Swarm Algorithms, Natural language processing, Recommender Systems, and Robotics (simulation).
Email: dadkhah@kntu.ac.ir

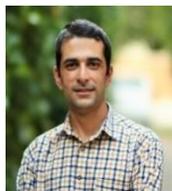

**Joobin Gharibshah** received the B.Sc. degree in software engineering, the master's degree in computer science, and the Ph.D. degree in computer science from the University of California Riverside, in 2020, USA.

His research interests include recommendation systems, search and ranking methods, natural language processing, and machine learning. He is an applied researcher at Ebay Inc.

Email: jghar002@ucr.edu